# Title: Observation of a Robust Zero-energy Bound State in Iron-based Superconductor Fe(Te,Se)

**Authors:** J.-X. Yin[1,2]*, Zheng Wu[2]*, J.-H. Wang[2], Z.-Y. Ye[1,2], Jing Gong[1], X.-Y. Hou[1], Lei Shan[1,3], Ang Li[3,2], X.-J. Liang[1], X.-X. Wu[1], Jian Li[2], C.-S. Ting[2], Z. Wang[5], J.-P. Hu[1,6], P.-H. Hor[2], H. Ding[1,3], S. H. Pan[2,1,3]†

**Affiliations:**

[1]Institute of Physics, Chinese Academy of Sciences, Beijing 100190, China.

[2]TCSUH and Department of Physics, University of Houston, Houston, Texas 77204, USA.

[3]Collaborative Innovation Center of Quantum Matter, Beijing, China

[4]Shanghai Institute of Microsystem and Information Technology, Chinese Academy of Sciences, Shanghai 200050, China.

[5]Department of Physics, Boston College, Chestnut Hill, Massachusetts 02467, USA

[6]Department of Physics, Purdue University, West Lafayette, Indiana 47907, USA.

†Corresponding author. E-mail: span@uh.edu.

*These authors contributed equally to this work.

**Abstract**: A robust zero-energy bound state (ZBS) in a superconductor, such as a Majorana or Andreev bound state, is often a consequence of non-trivial topological or symmetry related properties, and can provide indispensable information about the superconducting state. Here we use scanning tunneling microscopy/spectroscopy to demonstrate, on the atomic scale, that an isotropic ZBS emerges at the randomly distributed interstitial excess Fe sites in the superconducting Fe(Te,Se). This ZBS is localized with a short decay length of ~ 10 Å, and surprisingly robust against a magnetic field up to 8 Tesla, as well as perturbations by neighboring impurities. We find no natural explanation for the observation of such a robust zero-energy bound state, indicating a novel mechanism of impurities or an exotic pairing symmetry of the iron-based superconductivity.

**Main Text:** Superconductivity arises from the macroscopic quantum condensation of electron pairs. The symmetry of the wave-function of these pairs is one of the most essential aspects of the microscopic pairing mechanism. Since the impurity-induced local density of states (DOS) is sensitive to the pairing symmetry, it can be used to test the symmetry of the order parameter and to probe the microscopic pairing mechanism. Being a local probe with atomic resolution, scanning tunneling microscopy/spectroscopy (STM/S) (*1*) has played a key role in this respect, especially in the study of high-$T_C$ cuprate superconductors (*2*,*3*).

Since its discovery, new compounds of iron-based superconductor (IBSC) continue to be found. However, the pairing symmetry remains a central unresolved issue. So far,

STM/S studies regarding directly the local impurity scattering in single crystals of IBSCs have been limited to either weak scattering or unidentified impurities (*4-6*), despite the impurity-assisted quasi-particle interference experiments that support an unconventional pairing symmetry (*7, 8*). In our systematic experiments on various IBSCs, we have found that the excess Fe impurity atoms in Fe(Te,Se) induce strong local in-gap states. Since these excess iron impurities are known to suppress superconductivity efficiently (*9, 10*), a single crystal of Fe(Te, Se) containing a controlled amount of excess iron is a natural and promising system for the single atomic impurity experiments.

The as-grown Fe(Te,Se) single crystals usually contain a large amount of excess Fe that exist as single Fe atoms randomly situated at the interstitial sites in the crystal. These interstitial Fe impurity (IFI) atoms suppress the superconducting transition temperature $T_C$ at a rate of about 4K/1%Fe. The highest $T_C$ can be obtained when all the IFIs are removed. The removal of the IFIs in $Fe_{1+x}$(Te,Se) can be well controlled by an annealing process (*11*). In this experiment, we have studied single crystals with various IFI concentrations, from as-grown crystals (x = 1.8%) to crystals completely removed of IFIs (x = 0).

In Fig. 1A, we display an atomically resolved STM topographic image of a cleaved Fe(Te,Se) single crystal ($T_C$=14.5K), revealing the (Te,Se)-terminated surface, with the brighter spots being Te atoms and the less bright spots being Se atoms. On such a surface, at a temperature when the crystal is deep in the superconducting state, the tunneling spectrum exhibits a pair of sharp coherent peaks at ±1.5meV and a pair of side peaks at ±2.55meV, together with a stateless low-energy region, as shown in Fig. 1B. It is remarkable that the energy scales of the two pairs of peaks match well with the amplitudes of the two superconducting energy gaps on different Fermi surface (FS) sheets, as observed by angle resolved photoemission spectroscopy (ARPES) (*12*). These spectral features strongly suggest an *s*-wave-like (*7*), multi-gap nature of the superconducting state in Fe(Te,Se). The *s*-wave-like gaps are found to be rather spatially homogenous, as shown in Fig. 1C, thus providing us a clear background to 'view' the impurity states induced by an individual IFI.

In the image of a crystal with 0.5% IFI ($T_C$=12K), one can easily identify the IFI atoms showing as bright spots scattered on the exposed (Te,Se) surface, as displayed in Fig. 2A (*8, 13*). To pin-point the exact location of the IFIs, we zoom onto a single IFI (Fig. 2B), and find that it is located right at the center of the four neighboring Te/Se atoms on the surface (*14*). Spectroscopically, the individual IFI manifests itself as a sharp peak precisely at zero-bias, as shown in Fig. 2D, which is very different from the tunneling spectrum of the IFIs in the superconducting FeSe thin film (*15*). Furthermore, such a zero energy spectral peak is accompanied by a low-energy DOS depression without coherent peaks, which is in sharp contrast with the spectra taken far-away from the IFIs, suggesting a strong local scattering caused by the IFI and the consequential suppression of superconductivity.

To better reveal the characteristics of such local effects, we study a single IFI atom on a sample with a miniscule IFI content (x=0.1%, $T_C$=14K), as in Fig. 3A. A strong ZBS peak is observed in the tunneling spectrum taken at the center of the IFI site, as shown in Fig. 3B. The spatial pattern of this ZBS is rather circular, as seen from the zero-energy map in Fig. 3E, which is quite different from the cross-shape pattern of the Zn impurity in $Bi_2Sr_2Ca(Cu,Zn)_2O_{8+\delta}$ (*2*), indicating that the IFI scattering is quite isotropic. This can also be verified by the nearly identical spectra measured along a circle around the IFI (Fig. 3C). The ZBS peak is also very localized, only visible only within a region of ~ 10Å in diameter. A more quantitative analysis, i.e., fitting the zero-biased intensity along a line departing from the IFI (Fig. 3D), demonstrates an exponential decay of the ZBS intensity with a characteristic length of $\xi=3.5$Å (Fig. 3F), smaller than the typical coherent length of ~ 20 Å in the IBSCs (*16*). We also note that the ZBS peak remains strictly at zero-energy when measured away from center of the IFI site.

To confirm that the ZBS is indeed induced by IFI, we remove the IFI atom with the STM tip. As demonstrated in Figs. 3G and 3H, superconductivity is fully 'recovered' at the original site of the removed IFI, and the tunneling spectrum becomes identical to the spectra taken far away from the IFI. It is worthwhile to point out that the integrated low-energy spectral weight remains constant (fig. S2C), indicating that the spectral weight of the ZBS peak comes primarily from the superconducting coherent peaks. Moreover, the magnitude of the superconducting gap is seemingly unaffected by the IFI (Figs. 3C and D), suggesting that the IFI only weakens the superconducting phase coherence, but not the strength of the superconducting pairs.

Another aspect of this ZBS is its narrow line-width (~ 0.3 meV), which can be seen by removing the convolution effect of the finite temperature Fermi-Dirac distribution function of the normal state STM tip, as illustrated in Fig. 4A. This sharp ZBS peak broadens rapidly when temperature is raised, and vanishes completely at 15K (just above the bulk $T_C$) (Fig. 4A, inset), indicating the intimate relation between the ZBS and the bulk superconductivity.

Due to the breaking of time-reversal symmetry, an external magnetic field can cause a ZBS peak to split, such as the split of Andreev bound state observed in the cuprates (*17*). To our surprise, the ZBS peak observed on the IFIs remains unchanged against external magnetic fields. As demonstrated in Figs. 4B and 4C, the ZBS is not split, shifted, nor suppressed by a c-axis magnetic field up to 8T. In a typical spin-degenerate system ($g=2$), the Zeeman splitting is expected to be ~ 0.9 meV at 8T, which can be easily detected by our high-resolution STS. Furthermore, the ZBS peak remains at zero energy even when two IFI atoms are located near each other (~ 15 Å) as shown in Figs. 4D and F. This is quite unexpected, since two degenerated states will split when coupled quantum mechanically. However the ZBS peaks of the two closely located IFI atoms are found to be suppressed in intensity, possibly due to the local suppression of superconductivity or destructive interferences.

The explanation of the observed ZBS induced by individual IFI atoms appears to be a serious theoretical challenge based on the existing knowledge. Our observation of a *s*-wave-like full gap suggests the paring symmetry to be of the centrally debated $S^{++}$ or $S^{\pm}$ type (*18*). However, for either symmetry, the in-gap impurity bound states of a classical magnetic or non-magnetic impurity generally form a pair of spectral peaks, symmetrically respect to zero energy (*3, 20, 21*). These peaks will Zeeman-split when an external magnetic field is applied. Even for a quantum impurity, such as a Kondo impurity, the resonance peak is usually a bit off from zero-energy due to the existence of the finite potential scattering in the superconducting state and should exhibit Zeeman-like splitting as well (*1,22*). The only known conventional superconducting state that allows a ZBS is the *d*-wave pairing symmetry with an impurity at unitary limit (*1*, *2*). However, for this symmetry, quasi-particles tend to leak along the nodal directions, forming a four-fold-symmetry spectral pattern (*2*, *3*). It is evident that our observations of the isotropic ZBS induced by an IFI in the background of *s*-wave-like superconductivity are drastically inconsistent with the magnetic or non-magnetic impurity effects in either *d*- or *s*-wave superconductors.

The inaptness of the aforementioned conventional interpretations with all aspects of our experimental observations leads us to consider that the robust ZBS ought to be an indication of an unconventional pairing symmetry. We note that there are emerging theoretical proposals with new pairing symmetries for iron based superconductors, such as odd parity spin singlet pairing (*23*), tetrahedral and orbital pairing (*24*), concentric $s\pm$ (*25*), and *s+id* pairing (*26*). All of these proposals actually embrace the merits of both *s*- and *d*-wave order parameters by considering a special symmetry of the Fe-As/Se/Te bonding and indeed hint at a ZBS in a specific scattering channel. Besides these intrinsic exotic pairing proposals, it is also possible that exotic pairing is induced by IFI itself locally. For example, in a way similar to a diffusive ferromagnet (DF)/*s*-wave superconductor junction, an odd frequency spin triplet pairing may be induced at the interface, which can generate a ZBS (*27*). Nevertheless, it is largely unclear whether these exotic pairing symmetries are consistent with our observations. The robust ZBS, therefore, is a great challenge to the current theories and shall certainly help to advance our understanding of the superconducting state and its interaction with impurity atoms in general.

**Acknowledgments:**

The authors thank Zhong Fang, Xi Dai, Tao Xiang, D. -H. Lee, T. -K. Lee, G. -M. Zhang, Piers Coleman for stimulating discussions. We are grateful to the National Science Foundation and Ministry of Science and Technology of China and the Chinese Academy of Sciences for financial support.

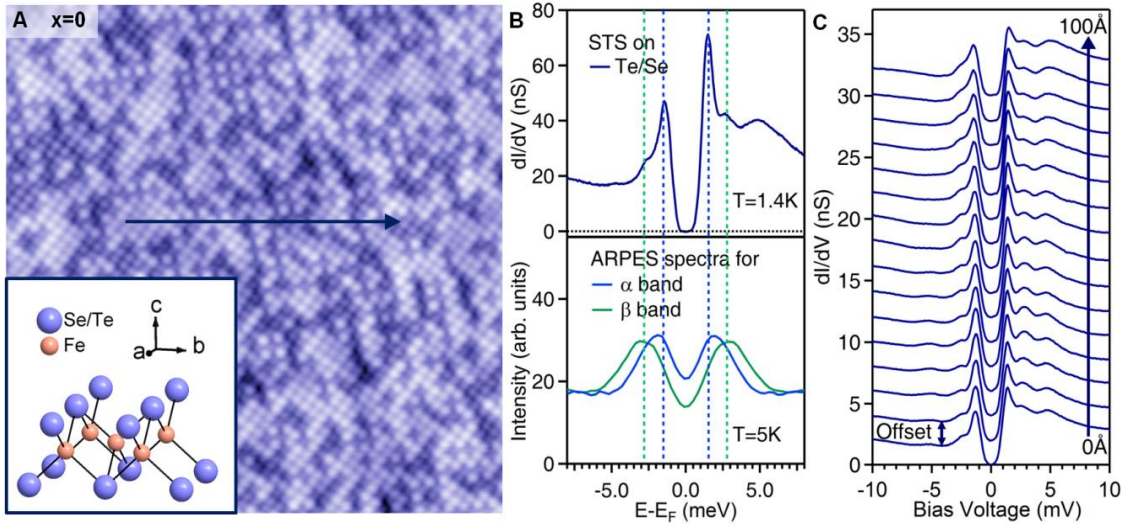

**Fig. 1**. Homogeneous two-gap structure on Fe(Te,Se). (**A**) Topographic image of (Te,Se) (180×180Å, T=1.5K, V=-100mV, I=0.1nA). Inset image is Fe(Te,Se) crystal structure. (**B**) Comparison between STS and ARPES data. The upper panel is dI/dV spectrum taken on (Te,Se) surface (T=1.4K, V=-10mV, I=0.5nA). The lower panel is symmetrized ARPES spectra (*12*). (**C**) STM spectra taken along the line shown on (**A**).

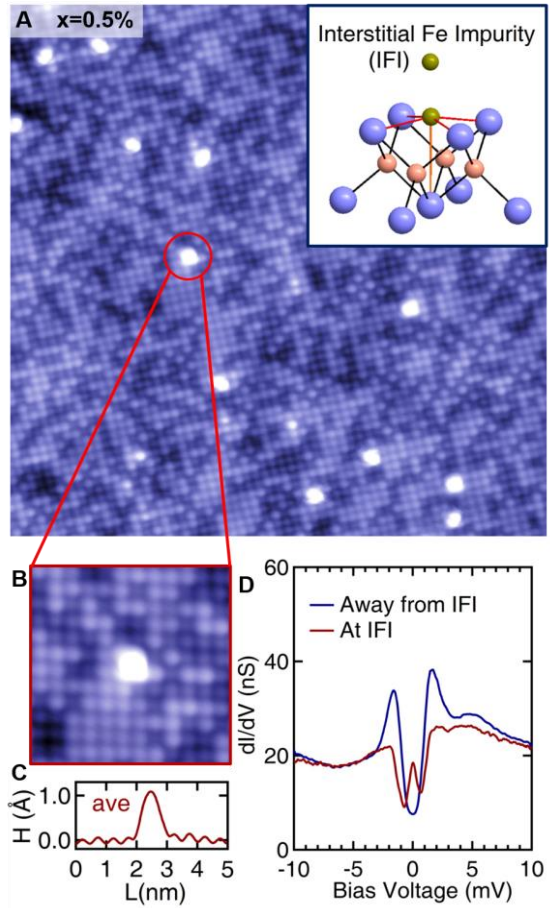

**Fig. 2.** Identification of IFI on $Fe_{1+x}$(Te,Se). (**A**) Topographic image of the (Te,Se) surface (200×200Å). Inset image is the crystal structure of $Fe_{1+x}$(Te,Se) with an IFI on the surface (Te,Se) layer (*14*). (**B**) A zoomed-in image showing an IFI surrounded by four 'bright' Te/Se atoms. (**C**) Averaged topographic profile over all the IFI atoms in (**A**). (**D**) Spectra taken at and away from IFI. All the data are acquired at 1.5K.

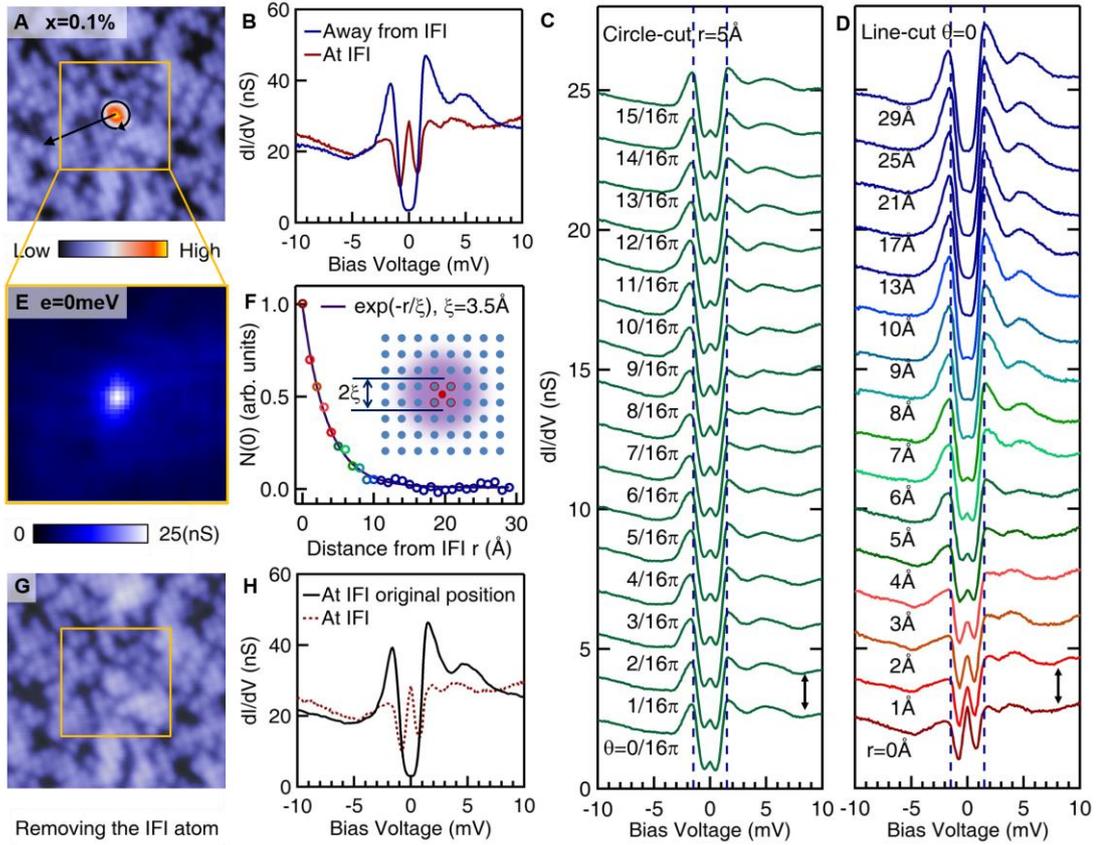

**Fig. 3.** Spatial evolution of the ZBS. (**A**) Topographic image of an isolated single IFI (100×100Å). (**B**) Spectra taken on top of and away from the IFI. (**C**) Spectra taken along the circle in (**A**) (r=5Å). (**D**) Spectra taken along the line in (**A**) (θ=0). (**E**) Zero-energy map for the box in (**A**). (**F**) Zero-energy peak value N(0) ( normalized to peak value ) versus distance from IFI r (data extracted from spectra in (**D**)). The solid curve is an exponential fit with ξ=3.5Å. Inset is the schematic image for spatial distribution of IFI scattering. (**G**) Topographic image of the same area as (**A**) with the IFI removed by STM tip. (**H**) Spectrum taken at the original position of the removed IFI. All the data are acquired at 1.5K.

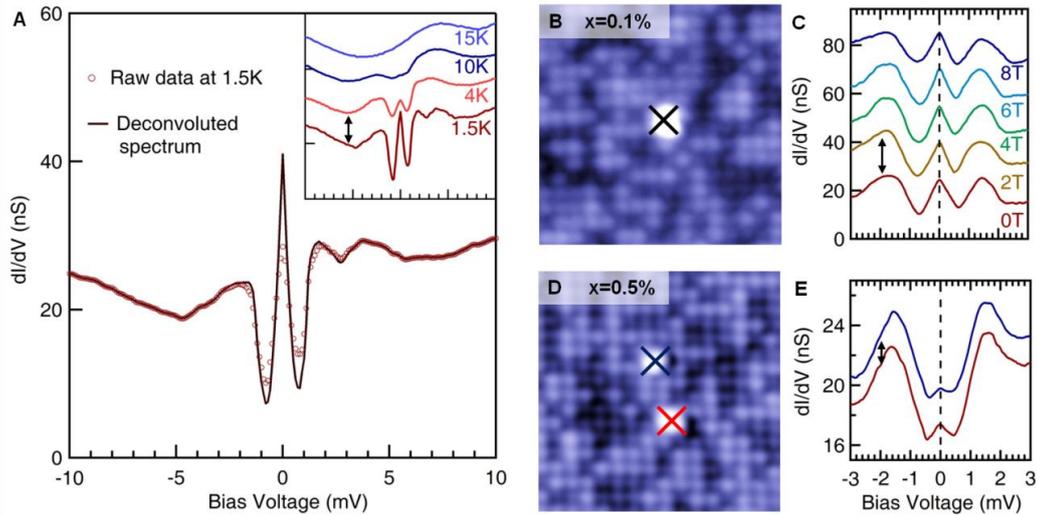

**Fig. 4.** Perturbing the ZBS by temperature, magnetic field and neighbouring IFI. (**A**) Circular markers represent the raw data taken at 1.5K. Solid curve represents the deconvoluted spectrum of the raw data by the Fermi-Dirac distribution function of 1.5K. Inset shows the spectra taken at the same IFI at different temperatures. (**B**) Topographic image (60×60Å, T=2.2K) of a single IFI. (**C**) ZBS spectra measured under different magnetic field along c-axis at IFI shown in (**B**) (T=2.2K). (**D**) Topographic image showing two IFI atoms are close in space (60×60Å, T=1.5K). (**E**) Spectra taken at the two IFI sites as shown in (**D**) (T=1.5K).

**Supplementary Materials:**

**Materials and Methods:**

All the crystals, with nominal composition FeTe0.5Se0.5, used in this study were grown using a self-flux method. A mixture of appropriate amount of iron (99.998 wt. %), tellurium (99.999 wt. %), and selenium (99.999 wt. %) powders was ground, pressed, and sealed in an evacuated double wall quartz ampoule after purging repeatedly with high purity Ar gas to ensure an oxygen free environment. The ampoule was then heated in a vertical tube furnace with the following procedure: Slowly heating over the temperature range of 400oC to 600oC, holding at 1100oC for 36 h, cooling in a rate of 1oC/h to 500oC, then quenching it into ice water. The phase purity of the crystals wass characterized using a Rigaku Geigerflex x-ray powder diffractometer on the powder of grounded crystals. Inductively coupled plasma mass spectrometry (ICP-MS) measurement of these samples gave Te:Se ratio=0.57:0.43, which might be due to the reaction of Se with quartz tube during sample growth. The content of interstitial Fe was tuned by post annealing process in evacuated quartz tubes. The DC magnetic susceptibility $4\pi\chi(T)$, measured using a Quantum Design Magnetic Properties Measurement System (MPMS), shows a sharp transition at onset of 14.5 K with transition width of 0.3 K and 1 K under zero-field cooled and field cooled process in fig. S1. The screening volume fraction is about 100% and Meissner fraction is as high as 53% after demagnetization factor correction, indicating high quality crystals with bulk superconductivity were used in our STM studies.

Samples were cleaved in cryogenic ultra-high vaccum at 4.2 K and were imaged with atomic resolution when inserted into STM head. STS spectra were recorded using a standard lock-in technique.

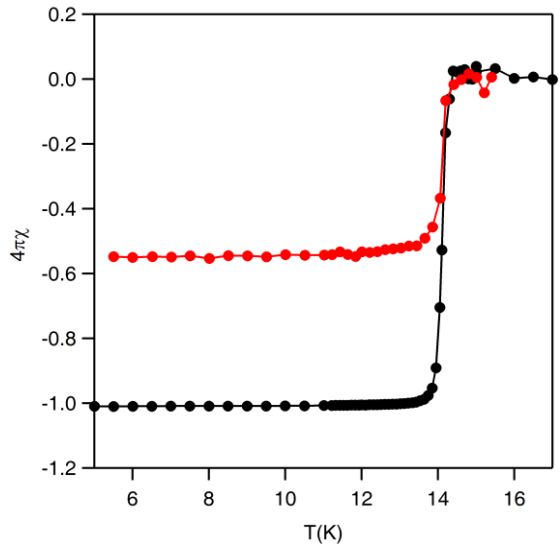

**Figure. S1.** DC volume susceptibility of Fe(Te, Se) measured under 1 Oe magnetic field. Black and red solid circle represent zero-field cooled and field cooled susceptibility, respectively.

**Further data analysis:**

Figure. S2A presents the full data of the line-cut spectra in Fig. 3A. Out of the scattering range (>10Å), the spectra show a homogeneous evolution as the case in pristine sample. Then we subtract each spectrum by a spectrum taken far away from the IFI and make their intensity plot in figs. S2B and D. It is clear that when moving toward the IFI atom, ZBS is increasing (marked by red) and superconducting gap peak value is decreasing (marked by blue). And the integrated spectra weight from -10meV to 10meV is almost invariant as shown in fig. S2C.

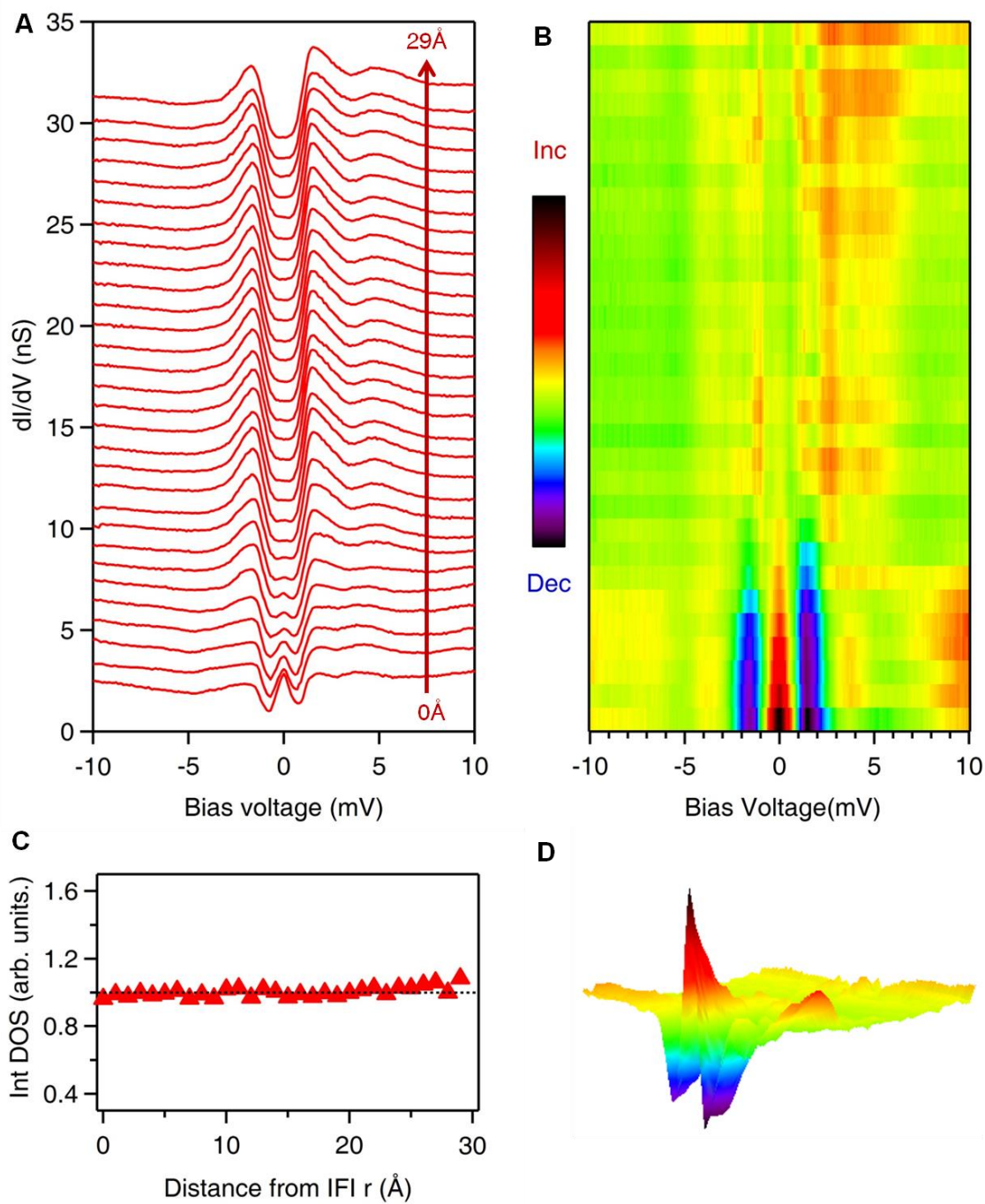

**Figure. S2.** Further analysis of the line-cut spectra. (**A**) line-cut spectra along the line in Fig. 3A. (**B**) Intensity plot of line cut spectra variation (subtracting each spectrum by a spectrum taken far away from IFI). Red color indicates increasing of the spectra value while blue color indicates decreasing of the spectra value. (**C**) Integrated density of states (Int DOS) from -10meV to 10meV for the line-cut spectra. (**D**) 3D plot of the data in (**B**).

**DFT calculation for interstitial Fe atom and discussion.**

Our DFT calculations employ the projector augmented wave (PAW) method encoded in Vienna ab initio simulation package (VASP), and the local density approximation (LDA) for the exchange correlation functional is used. Throughout this work, the cutoff energy of 400 eV is taken for expanding the wave functions into plane-wave basis. In the calculation, the Brillouin zone is sampled in the k space within Monkhorst-Pack scheme. The number of these k points are (5 ×5 ×1) in the surface calculation. We model a surface using 3 ×3 ×1 supercell containing two FeSe layers plus a vacuum layer of 15 Å. The bottom FeTe layer is frozen in the relaxation. The experimental lattice constants are adopted through the calculation. Forces are minimized to less than 0.02 eV/Å for slab calculation. The model in our calculation is shown in **Fig. S3a b**. The bond lengths of Fe1-Se1 and Fe1-Se2 are 2.79 and 2.60 Å, respectively. The tiny difference means that Fe1 can couple with both Se1 and Se2. To illustrate it, we plot the projected density of states of these atoms in **Fig. S4a** and **b**. p orbital of Se2 couples strongly with d orbital of Fe1 in the range from -5.0 eV to -3.0 eV. In contrast with Se2, there is an enhanced peak at -1.2 eV in the DOS of Se1, which is contributed by the hybridization of $d_z^2$ of Fe1 and $p_z$ of Se2. The DOS of Fe1 near Fermi level is greatly enhanced (with three sharp peaks) compared with that of Fe2 from **Fig. S4a**. The simulated STM images are shown in **Fig. S3c**, which are in agreement with that in experiment as shown in **Fig. 2**. We also calculated the charge density difference as shown in **Fig. S5**. There is some electron depletion near both up and down Se atoms, which clearly indicates that interstitial Fe forms covalent bonds with both of them. Especially, the depleted states are redistributed and mixed in the red area. This strong coupling between the IFI and both up and down Se layers may favour the interpretation of this ZBS with recent theoretical proposal of an odd parity spin singlet pairing (*23*) symmetry, which predicts a real space sign reversal of the up and down Se layers.

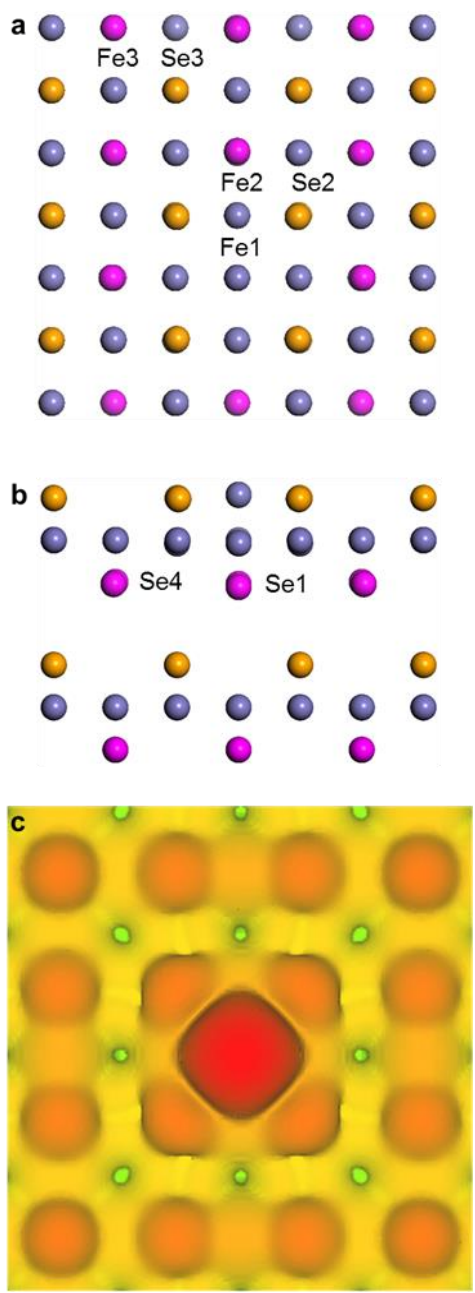

**Fig. S3 | DFT calculation on interstitial Fe in FeSe. a,b,** Geometric structure of Fe$_{1+x}$Se top view **a** and side view **b**, where interstitial Fe is at surface Se layer. **c**, The simulated STM image.

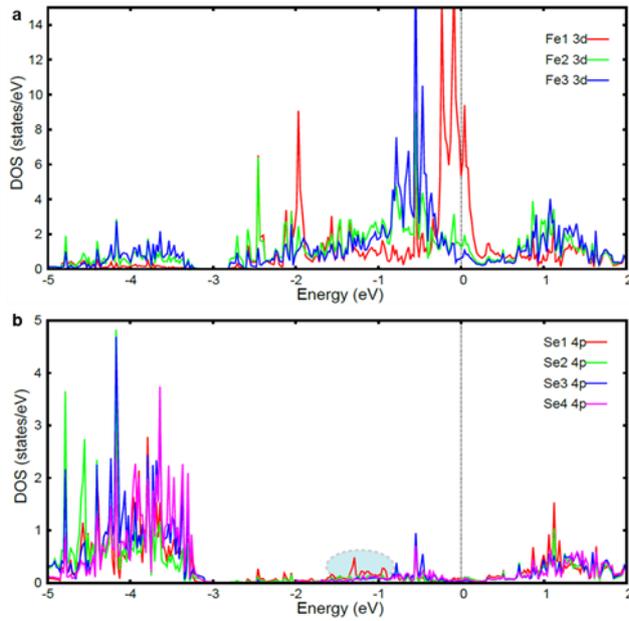

**Fig. S4 | DFT calculation on projected density of states (DOS) of special Fe and Se atoms in Fe$_{1+x}$Se. a,b**, DOS calculation for special Fe and Se atoms as marked in Fig. S3 a and b. The shaded area mark the strong coupling between the interstitial Fe atom the Se atom in the sublayer.

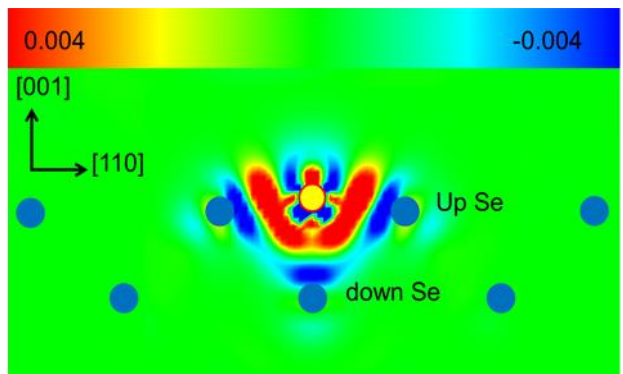

**Fig. S5 | DFT calculated charge density difference plot in units of e/(a.u.$^3$).**